\documentstyle[aps,prl,multicol,epsf]{revtex}
\begin{document}

\def\datestamp{September 10, 2003}

\title{Antiferromagnetism and phase separation 
in the $t$-$J$ model at low doping:\\
a variational study}

\author{D.~A.~Ivanov}
\address{
Paul Scherrer Institute, CH-5232 Villigen PSI, Switzerland}

\maketitle

\begin{abstract}
Using Gutzwiller-projected wave functions, I estimate the
ground-state energy of the $t$-$J$ model for several variational
states relevant for high-temperature cuprate superconductors.
The results indicate antiferromagnetism and phase separation
at low doping both in the superconducting state and in the
staggered-flux normal state proposed for the vortex cores.
While phase separation in the underdoped superconducting
state may be relevant for the stripe formation mechanism,
the results for the normal state suggest that similar 
charge inhomogeneities may also appear in
vortex cores up to relatively high doping values.
\end{abstract}

\begin{multicols}{2}

After many years of research, the issues of antiferromagnetism and
of phase separation in weakly-doped high-$T_c$ 
cuprates are far from being settled.
While it is generally accepted that short-range antiferromagnetic (AF)
correlations are crucial for the superconductivity in cuprates, it is not
clear whether the long-range antiferromagnetism at very low doping should be 
considered as a competing order on equal footing with the
superconductivity or simply as a minor side effect.
The latter point of view is implicitly
assumed in the resonating-valence-bond (RVB) scenario of
high-temperature superconductivity\cite{PWA}. In terms of ground-state
properties, the RVB approach may be recast into the language of variational
Gutzwiller-projected (GP) wave functions\cite{Gros1,Gros2,Yokoyama}. 
Recently a considerable progress has been reported in describing properties
of high-temperature superconductors with the help of GP wave
functions for the $t$--$J$ (or Hubbard) model\cite{Paramekanti}. 
Most studies of the GP wave functions neglect the long-range
AF ordering at low doping, and the resulting phase diagram contains
the superconducting phase starting from zero doping.
However, it is very easy to take AF order into account by explicitly
including it in the variational wave function. Within this
approach, the GP wave function for the $t$--$J$ model
is known to be energetically unstable with respect
to the AF order below the level of doping 
about 10\% (at $J/t=0.3$)\cite{Himeda}.

The phase-separation issue is a much more delicate subject than
antiferromagnetism: it is not decided even at the level of the $t$--$J$
model. While the phase separation at large $J/t$ is well established,
different studies do not agree about whether the phase separation
occurs in the physically relevant parameter range 
(at $J/t\sim 0.3$)\cite{EK-separation,Hellberg-Manousakis,%
Sorella-separation,Shih-separation}.

In this work I refine the numerical results of Ref.~\onlinecite{Himeda}
on the ground-state energy of the superconducting state in presence of
AF order and perform a similar analysis for the staggered-flux state recently
proposed to describe the normal state in the vortex cores\cite{Lee-vortex,%
Ivanov-Lee}. For both normal and superconducting states we find 
antiferromagnetism and phase separation at low doping. The phase
separation follows from the upward convexity
of the ground-state energy as a function of doping.
Within our approximation, the phase separation persists to higher
dopings than antiferromagnetism.
Consequently, the coexistence of AF order and superconductivity
is not realized as a homogeneous state (at least, for the
considered dimensionless parameter $t/J=3$). Instead, at
dopings lower than the phase-separation point $x_{\rm sep}$, it is
energetically favorable to split into two phases: the undoped
antiferromagnet (with the long-range AF order but without superconductivity)
and the superconductor with the doping $x_{\rm sep}$ 
(without the AF order). The relative areas of the two domains are fixed
by the average doping, and the actual shape of the domains is determined
by the Coulomb interaction at large distances (neglected at
the level of the $t$--$J$ model). This is the phase-separation
scenario of the stripe formation\cite{KE-stripes,Seibold,Han-Wang-Lee}.
Some numerical studies suggest an alternative point of view
that stripes appear already in the $t$--$J$ model without any long-range 
interactions\cite{White-Scalapino}. The results reported in this paper
do not have any implications on the latter scenario, since I do not
consider here any incommensurate spin-charge-density-wave states.
A priori it is possible that stripe states with energies lower than
the states studied here exist and may also be constructed
variationally by Gutzwiller projection.

Regarding the effect of the long-range Coulomb interaction,
I only remark that already nearest-neighbor repulsion may
act to reduce the phase separation. With increasing nearest-neighbor
repulsion, the phase-separation region shrinks to smaller doping
values and practically disappears at the repulsion strength
about 3--4$J$ --- approximately equal to the repulsion strength
required to suppress superconductivity. [Note that nearest-neighbor
repulsion may also favor stripe-like states not considered here.]

Finally, I comment on a similar phase-separation feature in the
staggered-flux normal state proposed recently to describe the
normal state in the vortex core \cite{Lee-vortex,Ivanov-Lee}. 
The phase separation in the normal state appears more prominent and
extends to higher doping values than in
the superconducting state. This
suggests that charge inhomogeneities (an analogue of stripes)
may also appear in the vortex core. However, as estimated
in Ref.\onlinecite{Ivanov-Lee}, the core size should be of order of
several lattice spacings, and large gradients of the order
parameter play an important role in determining
the structure of the core, together with the long-range
Coulomb interaction. Thus the problem of the vortex core
structure, even at the level of optimizing the 
Gutzwiller-projected wave function, becomes a very complicated one,
and our present analysis of uniform states is not sufficient for
solving it. 
We note here that insulating and antiferromagnetic regions in the
vortex cores have also been predicted in other approaches: in the
SO(5) theory\cite{Arovas} and in the unrestricted 
mean-field analysis\cite{Han-Lee}. 

In the rest of the paper, I present the details of our numeric analysis
of the Gutzwiller-projected wave functions.

The variational wave functions studied in this work are chosen to minimize
the energy of the $t$-$J$ Hamiltonian
\begin{equation}
 H = P_G \left[
 \sum_{ij} \left(-t c^{\dagger}_{i\alpha}c_{j\alpha} 
 + J (\vec{S}_i \vec{S}_j - {1\over4} n_i n_j) \right) \right] P_G \ ,
\label{t-J-hamiltonian}
\end{equation}
where the sum is taken over the pairs of
nearest-neighboring sites $i$ and $j$ on the two-dimensional
square lattice, $n_i$ denotes the hole density at a given site. 
Following the usual Gutzwiller-projection approach, 
we consider the variational wave functions of the form 
$\Psi_{\rm GP}=P_G \Psi_0$. Both here and in the Hamiltonian
(\ref{t-J-hamiltonian}) $P_G$ denotes the ``double'' projection: 
it projects out components with doubly occupied sites 
(the usual Gutzwiller projection) and further it also
fixes the total number of particles to the required value.
$\Psi_0$ is the ground state of the auxiliary (``mean-field'')
BCS Hamiltonian
\begin{eqnarray}
 H_{\rm BCS} &=& \sum_{ij} \left(-\chi_{ij} c^{\dagger}_{i\alpha}c_{j\alpha} 
 + \Delta_{ij} (c^{\dagger}_{i\uparrow} c^{\dagger}_{j\downarrow} -
 c^{\dagger}_{i\downarrow} c^{\dagger}_{j\uparrow}) + {\rm h.c.} \right)
\nonumber \\
&+& \sum_i (-1)^i h \sigma_i^z \ ,
\label{BCS-hamiltonian}
\end{eqnarray}
where $\sigma^z_i=
c^{\dagger}_{i\uparrow} c_{i\uparrow}-
c^{\dagger}_{i\downarrow} c_{i\downarrow}$ is the $z$-magnetization
at the site $i$. As usual, we take the nearest-neighbor $d$-wave
ansatz: $\chi_{ij}=\chi$ on nearest-neighbor links, $\chi_{ii}=\mu$
is the on-site chemical potential, $\Delta_{ij}=\pm\Delta$,
with $\pm$ for vertical and horizontal nearest-neighbor
links respectively. Within this variational ansatz, the wave function
depends on the three dimensionless parameters: $\Delta/\chi$,
$\mu/\chi$, and $h/\chi$. The parameter $h$ represents an
artificial staggered magnetic field acting on spins to produce
the long-range AF order. At zero value of $h$, the minimization
has been performed by many 
authors\cite{Gros2,Yokoyama,Paramekanti,Ivanov-Lee}. We further adopt the
numbers reported in our earlier publication Ref.\onlinecite{Ivanov-Lee}.
As in that work, we compute the ground-state energy by the
variational Monte Carlo method (see e.g.\ Ref.\cite{Gros2} for details
of the method), on the square lattice 22$\times$22 with the boundary
conditions periodic along one direction and antiperiodic along the
other direction. For this system size, the finite-size corrections
to the ground-state energy are estimated to be of the order of the
Monte Carlo statistical errors (about $10^{-3}J$) [in the staggered-flux
state the finite-size corrections are somewhat bigger because of
the discretization of the Fermi pockets].
The parameter of the $t$-$J$ model is taken to be
$t/J=3$ throughout the paper.

To check for an instability with respect to the AF ordering, we
further minimize the variational energy as a function of $h$
while keeping $\Delta/\chi$ and $\mu/\chi$ constant (in principle,
the minimization should be performed by varying all the three parameters 
simultaneously; we can however check that, in the vicinity of the
energy minimum, the errors from our simplified minimization procedure 
are negligible). Instead of characterizing the AF state by the
fictitious field $h$, we use a physically significant quantity, 
the staggered magnetization.
In Fig.~1 we plot the staggered magnetization in the optimal wave
function as a function of doping $x$. We find the instability
towards the AF ordering below $x_{\rm AF} =0.11$. 
So far the computations repeat those in Ref.~\onlinecite{Himeda}.
Thus obtained values of staggered magnetization should only be considered
as an indication of an instability, and not as good numerical estimates
of the actual staggered magnetization in the ground state: for example,
in the undoped case, the variational estimate is much higher than the
known exact value\cite{Heisenberg}.


\begin{figure}
\epsfxsize=0.5\hsize
\centerline{\epsffile{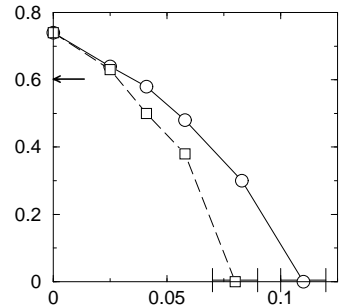}}
\bigskip
\caption{Staggered magnetization $M=\langle (-1)^i \sigma^z_i \rangle$
as a function of doping at $t/J=3$ for the superconducting (circles)
and staggered-flux (squares) states. Vertical error bars are of the
order of the symbol size. Error bars of the phase transition points
are about 0.01 (shown). Our results for the superconducting state
agree with those in Fig.~2 of Ref.~\protect\onlinecite{Himeda}.
The arrow shows the exact value ($M=0.60$) of the staggered magnetization 
in the undoped Heisenberg antiferromagnet.}
\label{fig1}
\end{figure}


If we analyze the resulting ground-state energy as a function
of doping, we observe that it is upward convex at low doping. This
implies that at those dopings phase separation occurs. In Fig.~2 we
plot the ground-state energy with the subtracted linear part
$E-\mu_{\rm sep} x$ as a function of doping. The slope $\mu_{\rm sep}$
of the limiting tangent line gives the chemical potential
at the phase separation point. The point of contact determines the
critical doping $x_{\rm sep}$ below which the phase separation occurs.
At $t/J=3$, we estimate $x_{\rm sep}=0.13$ and $\mu_{\rm sep}=-5.65 J$
(this value of $x_{\rm sep}$ agrees with the estimates of 
Ref.\onlinecite{Hellberg-Manousakis}).

Note that for the $t$--$J$ model with $t/J=3$ we obtain
$x_{\rm sep}>x_{\rm AF}$, i.e.\ antiferromagnetism
does not mix with superconductivity in a uniform phase;
instead, phase separation occurs between the undoped antiferromagnet
and the superconductor at doping $x_{\rm sep}$ without
antiferromagnetism. However, this inequality appears to be 
non-universal, and could, in principle, be reversed 
by adding other terms like
next-nearest-neighbor hopping or nearest-neighbor repulsion
(see the discussion of nearest-neighbor repulsion below).

We may further perform the same steps as above for the ``normal''
staggered-flux (SF) state proposed to describe vortex cores
in the mixed state\cite{Lee-vortex,Ivanov-Lee}. 
We first test the SF state for the
AF instability and find that, similarly to the superconducting state,
the SF state favors AF order at very low doping. We find that the
corresponding staggered magnetization (also plotted in Fig.~1) is smaller
than in the superconducting state at the same doping. The critical
value of doping $x_{\rm AF}^{(SF)}$ is also smaller than in the
superconducting state (we estimate $x_{\rm AF}^{(SF)}=0.08$).
Thus antiferromagnetism in vortex cores appears at the first glance
more fragile than in the bulk of the superconductor.


\begin{figure}
\epsfxsize=1.0\hsize
\centerline{\epsffile{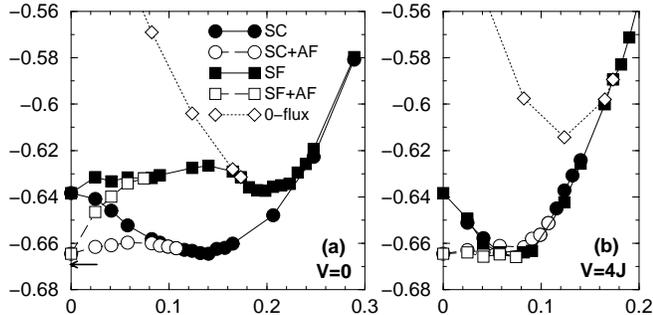}}
\bigskip
\caption{Energies of different GP wave functions with linear part
subtracted ($E-\mu_{\rm sep}x$) as a function of doping $x$.
The five different variational wave functions are compared
(optimized in energy within each class): superconducting
without antiferromagnetism (solid circles, solid line), 
superconducting with antiferromagnetism (empty circles, dashed line),
staggered-flux without antiferromagnetism (solid squares, solid line),
staggered-flux with antiferromagnetism (empty squares, dashed line),
and zero-flux (projected Fermi sea, empty diamonds, dotted line).
Left panel {\bf (a)} shows energies in the absence of NN repulsion
($V=0$), with $\mu_{\rm sep}=-5.65$. Right panel {\bf (b)} corresponds
to $V=4J$, with $\mu_{\rm sep}=-4.85$ [NN repulsion is included at the
perturbation level via (\ref{nn-perturbation})]. The energies are
in the units of $J$ per lattice site. Error bars are smaller than
the symbol size. For comparison, the arrow in panel (a) shows the
exact ground-state energy in the undoped case 
($E=-0.669J$)\protect\cite{Heisenberg}.
}
\label{fig2}
\end{figure}


However the phase-separation effect in the SF state is more
pronounced than in the superconducting state: we estimate
$x_{\rm AF}^{(SF)}=0.21$ --- far in the ``overdoped''
region of the phase diagram. We summarize our results on the ground-state
energies of different states in Fig.~2 (the difference between the
energies of the superconducting and SF states was reported previously
in Ref.\onlinecite{Ivanov-Lee} as the condensation energy for the
superconducting state). Our results on the phase separation in the
SF state indicate that charge inhomogeneities (similar to stripes
in underdoped cuprates) are likely to form in the vortex cores up
to rather high doping values. As in the case with stripes, the
actual structure of inhomogeneities should be determined by the
long-range Coulomb interactions. Such inhomogeneities may play
a role in producing the ``subgap state'' features observed in the
density of states in the STM experiments\cite{Pan-Renner}.

As a simplified version of long-range Coulomb interaction, we
consider the effect of nearest-neighbor (NN) repulsion on the
phase separation. The NN repulsion is included as the
additional term in the Hamiltonian (\ref{t-J-hamiltonian}),
$H_V=V\sum_{ij} n_i n_j$, where the sum is taken over all
NN pairs of sites [$n_i$ denotes the hole density, as before;
$V$ is the repulsion strength in addition to the attraction
$-J/4$ already present in the Hamiltonian (\ref{t-J-hamiltonian})].
If we neglect interaction between holes, $H_V$ might be thought to
scale as $x^2$ with doping $x$. Then a small repulsion energy
$V$ might be already sufficient to change the convexity
of the total energy as a function of doping and hence to
suppress phase separation. In reality, in the superconducting
antiferromagnetic state at small doping, the
attraction between holes is quite strong, and scaling of $H_V$ with
doping is closer to $x$ than to $x^2$. In Fig.~3a,b we plot the
nearest-neighbor correlation function $\langle n_i n_j \rangle$
normalized by $x$ and by $x^2$, respectively.


\begin{figure}
\epsfxsize=1.0\hsize
\centerline{\epsffile{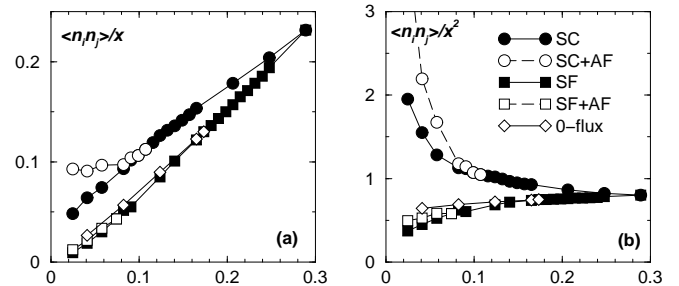}}
\bigskip
\caption{Nearest-neighbor correlation of the hole density
$\langle n_i n_j \rangle$ as a function of doping for the same
wave functions as in Fig.~\ref{fig2}. Panel {\bf (a)} shows
nearest-neighbor correlation divided by the average hole density
$x$, and in panel {\bf (b)} the same correlation is divided by
$x^2$.}
\label{fig3}
\end{figure}


To estimate the effect of NN repulsion, we approximate
the ground-state energy with repulsion as
\begin{equation}
E(x)=E_0(x)+2V \langle n_i n_j \rangle\, ,
\label{nn-perturbation}
\end{equation}
where $E_0(x)$ is the variational ground-state energy without repulsion,
and the average is taken over the variational ground state without
repulsion. In this approximation, we can estimate the repulsion
strength necessary to reverse the convexity of the ground-state
energy $E(x)$. With increasing the NN repulsion strength $V$,
the upward convexity of the ground-state energy diminishes, and
the phase-separation point $x_{\rm sep}$ shifts towards smaller
doping (both in the superconducting and the SF states). At the
same time, the superconductivity gets also strongly reduced
(at least in its present nearest-neighbor $d$-wave ansatz), with
the superconducting transition point shifting to smaller doping.
The antiferromagnetism at finite doping is also reduced, 
but not so strongly as the superconductivity.

With increasing the NN repulsion strength $V$,  
the phase separation eventually becomes undetectable within our
numerical errors at $V \sim 3..4J$.
At the same time, at the repulsion strength of about 4$J$,
our superconducting wave function looses the energy
competition to the staggered-flux one: the price of nearest-neighbor
holes overweighs the energy gain from the superconductivity.
It is possible however that at such strong NN repulsion another
superconducting ansatz (e.g., involving next-nearest-neighbor
pairing) would be more energetically favorable (search for such
states goes beyond the scope of the present paper). We illustrate
the effect of NN repulsion in Fig.~2b where the marginal situation
is shown: the superconducting state is nearly equal in energy
to the SF state at $V=4J$ [with NN repulsion taken into account
only to the lowest order in the perturbation theory (\ref{nn-perturbation})].
Note that our numerical results show a rather flat doping dependence
of the ground-state energy at very low doping, and this leaves
unresolved the issue of whether phase separation at infinitesimally small
doping may survive arbitrarily strong NN repulsion.
Of course, the above treatment of the hole interaction is only a 
rough estimate. As suggested in Ref.~\onlinecite{Sorella-Lanczos},
the variational wave function may be further improved by including
additional Jastrow-type factors. Such a modification of the wave
function apparently affects the hole-hole correlations\cite{Sorella-Lanczos},
and may therefore be important for a proper assessment of the
effects of the hole-hole interaction.

The variational study reported in this paper should be taken with care
when applied to actual cuprate superconductors:
the phase-separation effect in the superconducting state
is weak, and many additional factors may change
the picture outlined here. However, I believe that some qualitative
conclusions may withstand small perturbations of the model:
(1) staggered-flux phase is more disposed to phase separation than the
superconducting state. This may produce charge inhomogeneities 
inside the vortex cores; (2) antiferromagnetic ordering at low doping
enhances phase separation (this may be a general property of phase
separation in doped Mott insulators, see, e.g., 
Ref.~\onlinecite{Lee-Kivelson} and references therein); 
(3) both in the staggered-flux state and
in the superconducting state, the long-range antiferromagnetic ordering 
occurs only at low doping (below $\sim$0.1). Thus static AF order in
vortex cores\cite{Kakuyanagi}, if confirmed, possibly
indicates regions with reduced hole concentration.

I thank P.~A.~Lee for many discussions and comments on the
manuscript. I am grateful to  F.~Becca for helpful comments. 
The computations were performed on the Beowulf cluster Asgard at ETH Z\"urich.

\end{multicols}
\end{document}